\documentclass[smallextended]{svjour3}       % onecolumn (second format)
\smartqed  % flush right qed marks, e.g. at end of proofs

\usepackage{graphicx,psfrag,color}% Include figure files
\usepackage{dcolumn}% Align table columns on decimal point
\usepackage{bm}% bold math
\usepackage{mathbbol}
\usepackage{amssymb,amsmath,amsfonts,amssymb}

\begin{document}

%\begin{frontmatter}

\title{Robust and efficient transport of two-qubit entanglement via disordered spin chains}
\titlerunning{Robust and efficient transport of entanglement via disordered spin chains}

\author{Rafael Vieira  \and Gustavo Rigolin}


\institute{Rafael Vieira \at
           Departamento de F\'{i}sica,
	   Universidade Federal de S\~ao Carlos, S\~ao Carlos, SP 13565-905,
	   Brazil          %  \\
          \and
           Gustavo Rigolin \at
           Departamento de F\'{i}sica,
	   Universidade Federal de S\~ao Carlos, S\~ao Carlos, SP 13565-905,
	   Brazil \\  
           \email{rigolin@ufscar.br}  
}

\date{Received: date / Accepted: date}

\maketitle

\begin{abstract}
We investigate how robust is the modified XX spin-1/2 chain of %R. Vieira and G. Rigolin (2018)
[R. Vieira and G. Rigolin, Phys. Lett. A \textbf{382}, 2586 (2018)]
in transmitting entanglement when several types of disorder 
and noise are present. 
First, we consider how deviations 
about the optimal settings that lead to almost perfect transmission of a maximally entangled two-qubit state
affect the entanglement reaching the other side of the chain. 
Those deviations are modeled by static, dynamic, and fluctuating disorder. 
We then study how spurious or undesired interactions and external magnetic fields diminish 
the entanglement transmitted through the chain. For chains of the order of hundreds of qubits,
we show for all types of disorder and
noise here studied that the system is not appreciably affected when we have weak disorder
(deviations of less than $1\%$ about the optimal settings) and that for moderate disorder  
it still beats the standard and ordered XX model when deployed to accomplish
the same task.
\end{abstract}

%\pacs{03.65.Ud}{Quantum entanglement}
%\pacs{03.67.Bg}{Entanglement production}
%\pacs{03.67.Hk}{Quantum communication}

\section{Introduction} 

An important tool in the practical implementation of quantum computation and communication tasks is
the reliable transmission of quantum states from one location to another \cite{ben00}.  
Quantum information, or equivalently a quantum state, can be sent from one party (Alice) to another (Bob)
in at least three ways, namely, via direct transmission, via the quantum teleportation protocol \cite{ben93},
and via
spin chains \cite{bos03}. This last strategy, where Alice and Bob are connected by a spin chain through which 
they can send quantum states to each other is the main focus of this work.
This task is accomplished by properly tuning the interaction among the qubits in the chain such that after 
the system evolves a certain time $t>0$, a state prepared by one party at $t=0$ reaches 
the other one with high fidelity. Note that here as well as in the quantum teleportation protocol no physical
system is transmitted from Alice to Bob. 

One important characteristic of a spin chain when employed to transmit quantum states 
is that the interaction strength among the qubits, once set up to give a high fidelity transmission,
is fixed along the time evolution of the system. Also, by using spin chains as the mean through which 
quantum information is sent back and forth within a silicon-based quantum chip, we will deal with the same
physical
system in order to process and transmit quantum information. In this way we avoid the technologically difficult 
task of integrating different physical platforms, one for processing and other for
transmitting information \cite{bos03}.

The several works investigating the usefulness of spin chains to perform 
quantum communication tasks have centered their focus on one of the 
three following scenarios, namely, the transmission of a single qubit 
\cite{bos03,chr04,nik04,sub04,osb04,chr05,woj05,kar05,har06,huo08,gua08,ban10,kur11,god12,apo12,lor13,hor14,shi15,pou15,zha16,che16,est17}, 
the generation of entanglement between the two ends or two specific sites of the chain 
\cite{bos03,chr05,li05,har06,ban10,lor13,est17b,apo18}, and the transmission of 
quantum states composed of two or more qubits \cite{nik04,sub04,chr05,shi05,lor13,sou14,lor15,vie18}.
We should also mention important contributions on the usefulness of chains of continuous variable systems in transmitting quantum
states \cite{cir97,ple04,sem05,har06,nic16}. Most of the aforementioned works studied strictly one dimensional chains
with open or periodic boundary conditions and extensive investigations on the performance of those systems in the presence of disorder and
noise can be found in \cite{nik04,chi05,fit05,bur05,pet10,zwi11,bru12,nik13,kur14,ash15,pav16,ron16,lyr17,lyr17a,lyr17b}. 

\begin{figure}[!ht] 
\begin{center}
%\onefigure{fig1.eps}
\includegraphics[width=11.5cm]{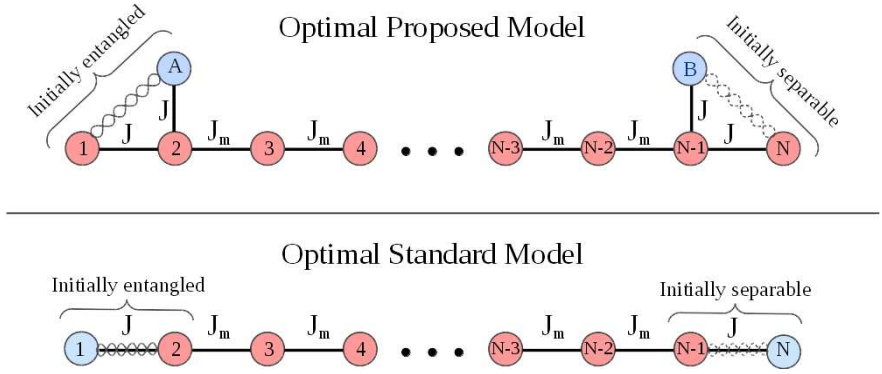}
\caption{%(color online) 
Upper panel: Alice's qubits $A$ and $1$ are set at $t=0$ as the maximally entangled Bell state $|\Psi^+\rangle=(|01\rangle + |10\rangle)/\sqrt{2}$
while the remaining qubits are set in the state $|0\rangle$. The system then evolves according to the Hamiltonian $H$, Eq.~(\ref{ham0}), 
and at a specific time $t>0$ Bob's qubits $N$ and $B$ become highly entangled. 
For the ordered and noiseless case Bob's qubits become an almost perfect replica of Alice's qubits if 
$J$ and $J_m$ are properly adjusted \cite{vie18}.
Lower panel: Alice's qubits $1$ and $2$ are initially $|\Psi^+\rangle$
and the remaining ones are in the state $|0\rangle$.  The system then evolves according to the standard XX Hamiltonian 
and at a specific time $t>0$ Bob's qubits $N-1$ and $N$ become entangled. However, the performance for the standard model decreases considerably 
when we add more and more spins to the chain \cite{vie18}. This is not the case with the proposed model, where one can always find for any size
of the chain a pair of values $J$ and $J_m$ such that the entanglement reaching Bob is for all practical purposes
the maximum value possible, i.e., Bob always gets the Bell state $|\Psi^+\rangle$ \cite{vie18}.
}
\label{fig1}
\end{center} 
\end{figure}

In this work our main goal is to investigate the performance of the quantum state transfer
protocol presented in \cite{vie18} when subjected to several types of noise and disorder. 
The protocol of \cite{vie18} was specifically built to 
transmit maximally entangled two-qubit states from Alice to Bob and to be simple in 
its construction and operation, namely, we avoided any modulation in the coupling constants between the 
qubits along the chain \cite{chr04,nik04,kay05} and we excluded the use of external magnetic fields to drive the transmission of 
the quantum state from Alice to Bob \cite{shi05,kay11,lor13,lor15}.
With such stringent requirements, we could achieve almost perfect entanglement transmission by going 
slightly beyond a one dimensional spin chain. In Fig. \ref{fig1} we show the optimal configurations for the model
proposed in \cite{vie18} and for the standard strictly one dimensional system usually seen in the literature.
We also point out a related geometry given in the work of Chen \textit{et al.} \cite{che16} aiming at sending
a single qubit from $1$ to $N$\!. 

\section{The model}
\label{tools}

The ordered system is described by the following slightly modified XX Hamiltonian,  
\begin{equation}
H = H_A + H_M + H_B,
\label{ham0}
\end{equation}
where
\begin{eqnarray}
H_A  \hspace{-.02cm}&=&\hspace{-.02cm} J (\sigma_1^x\sigma_2^x+\sigma_1^y\sigma_2^y) +J (\sigma_A^x\sigma_2^x+\sigma_A^y\sigma_2^y), \nonumber \\
H_M  \hspace{-.02cm}&=&\hspace{-.02cm}  \sum_{j=2}^{N-2}J_m(\sigma_j^x\sigma_{j+1}^x+\sigma_j^y\sigma_{j+1}^y), \nonumber \\
H_B  \hspace{-.02cm}&=&\hspace{-.02cm} J (\sigma_{N-1}^x\sigma_N^x\!+\!\sigma_{N-1}^y\sigma_N^y) \!+\! 
J (\sigma_{N-1}^x\sigma_B^x\!+\!\sigma_{N-1}^y\sigma_B^y). \nonumber \\
\label{ham}
\end{eqnarray}

The notation $\sigma_i^\alpha\sigma_j^\alpha$ means $\sigma_i^\alpha \otimes \sigma_j^\alpha$, the superscript
indicates a specific Pauli matrix, and the subscript denotes the qubit acted by it. Here   
$\sigma^z|0\rangle=|0\rangle,\sigma^z|1\rangle=-|1\rangle,
\sigma^x|0\rangle=|1\rangle,\sigma^x|1\rangle=|0\rangle,\sigma^y|0\rangle=i|1\rangle,\sigma^y|1\rangle=-i|0\rangle$,  and $i=\sqrt{-1}$. 
We can get the standard XX model setting $J=0$ and letting the summation for $H_M$
run from $1$ to $N-1$. In Ref. \cite{vie18} it was shown that for a system of $N+2$ qubits, with $N=100$, Bob gets
about $99\%$ of the entanglement created by Alice if $J_m/J=49.98$ in the proposed model.
And for small values of $J_m/J$, namely, $J_m/J\leq 5$, we get $82\%$ of the entanglement created by Alice 
arriving at Bob if $J_m/J=2.86$. The strictly linear chain (standard model with $N=100$ qubits) transmits only about 
$40\%$ of the entanglement with Alice in the best scenario (when we set $J_m/J=2.32$).

In order to study how disorder and noise affect the proposed model, we work with the following Hamiltonian,
\begin{equation}
\tilde{H} = \tilde{H}_A + \tilde{H}_M + \tilde{H}_B +\tilde{H}_z+\tilde{H}_{zz},
\label{ham01}
\end{equation}
where
\begin{eqnarray}
\tilde{H}_A  \hspace{-.02cm}&=&\hspace{-.02cm} J_{1,2}(t) (\sigma_1^x\sigma_2^x+\sigma_1^y\sigma_2^y) +J_{A,2}(t)(\sigma_A^x\sigma_2^x+\sigma_A^y\sigma_2^y), \nonumber \\
\tilde{H}_M  \hspace{-.02cm}&=&\hspace{-.02cm}  \sum_{j=2}^{N-2}J_{j,j+1}(t)(\sigma_j^x\sigma_{j+1}^x+\sigma_j^y\sigma_{j+1}^y), \nonumber \\
\tilde{H}_B  \hspace{-.02cm}&=&\hspace{-.02cm} J_{N-1,N}(t) (\sigma_{N-1}^x\sigma_N^x\!+\!\sigma_{N-1}^y\sigma_N^y) \!+\! 
J_{N-1,B}(t) (\sigma_{N-1}^x\sigma_B^x\!+\!\sigma_{N-1}^y\sigma_B^y). \nonumber \\
\tilde{H}_z  \hspace{-.02cm}&=&\hspace{-.02cm}  h_{A}(t)(\mathbb{1}-\sigma_A^z) + \sum_{j=1}^{N}h_{j}(t)(\mathbb{1}-\sigma_j^z) + 
h_{B}(t)(\mathbb{1}-\sigma_B^z), \nonumber \\
\tilde{H}_{zz}  \hspace{-.02cm}&=&\hspace{-.02cm}  \Delta_{A,2}(t)\sigma_A^z\sigma_{2}^z + \sum_{j=1}^{N}\Delta_{j,j+1}(t)\sigma_j^z\sigma_{j+1}^z 
+  \Delta_{N-1,B}(t)\sigma_{N-1}^z\sigma_{B}^z. 
\label{ham1}
\end{eqnarray}
Here $\tilde{H}_A,\tilde{H}_M,$ and $\tilde{H}_B$ extend the proposed model such that 
the coupling constants $J_{i,j}(t)$ may now change with time and with position (qubit's lattice location), allowing
us to model several types of disorder as later explained.
$\tilde{H}_z$ and $\tilde{H}_{zz}$ represent, respectively, external magnetic fields and 
the $\sigma_j^z\sigma_{j+1}^z$ interaction, which are two types of noise that may act on the proposed model.
When $J_{A,2}(t)=J_{1,2}(t)=J_{N-1,N}(t)=J_{N-1,B}(t)=J$, $J_{j,j+1}(t)=J_m$, for $j=2,\ldots,N-2$,
and all $h_j(t)$ and $\Delta_{i,j}(t)$ are zero, we recover the proposed model of Ref. \cite{vie18}.

Hamiltonian (\ref{ham01}) is such that $[H,Z]=0$, i.e., it commutes with the operator $Z=\sigma_A^z+ \sum_{j=1}^{N}\sigma_j^z+\sigma_B^z$. 
This implies that the number of spins up and down (excitations) does not change in time, which restricts considerably the size of 
the Hilbert space needed to describe the system. In this scenario, the Schr\"odinger equation can 
be efficiently solved when the system has just a few excitations \cite{vie18}.   

Following Ref. \cite{vie18}, we aim at sending from Alice to Bob entangled states containing just one excitation. 
In this case any system of $N+2$ qubits, as depicted in Fig. \ref{fig1},
is described by the linear combination of $N+2$ single-excitation states, 
\begin{equation}
|\Psi(t)\rangle = \sum_j c_{\!_j}(t)|1_j\rangle,
\label{psit}
\end{equation}
where $j=A,1,2, \ldots, N-1,N,B$, and
%
%\begin{equation}
%|1_j\rangle = \sigma_j^x|000\cdots\hspace{-.5cm} \underbrace{0}_\text{j-th qubit}\hspace{-.5cm} \cdots 000\rangle 
% = |000\cdots\hspace{-.5cm} \underbrace{1}_\text{j-th qubit}\hspace{-.5cm} \cdots 000\rangle. 
%\end{equation}
\begin{equation}
|1_j\rangle = \sigma_j^x|00\cdots\hspace{-.25cm} \underbrace{0}_\text{j-th qubit}\hspace{-.25cm} \cdots 00\rangle 
 = |00\cdots\hspace{-.25cm} \underbrace{1}_\text{j-th qubit}\hspace{-.25cm} \cdots 00\rangle. 
\end{equation}

We assume that when $t=0$ Alice's qubits $1$ and $A$ are the maximally entangled 
Bell state $|\Psi^+\rangle=(|01\rangle+|10\rangle)/\sqrt{2}$, leading to the following 
initial state for the whole system, $|\Psi(0)\rangle=(|010\cdots 0\rangle+|10\cdots 0\rangle)/\sqrt{2}$. 
In the notation of Eq.~(\ref{psit}) we have at $t=0$ 
\begin{eqnarray}
\hspace{-.02cm}c_{\!_A}(0)=c_{\,_1}(0)=1/\sqrt{2} & \hspace{.04cm}\mbox{and} & 
\hspace{.04cm}c_{\!_j}(0)=0, \hspace{.1cm} \mbox{for} \hspace{.1cm} j \neq A,1.
\label{initial1}
\end{eqnarray}

After substituting Eq.~(\ref{psit}) into the Schr\"odinger equation 
$$
i\hbar \frac{d|\Psi(t)\rangle}{dt} = \tilde{H}(t) |\Psi(t)\rangle
$$ 
and left multiplying it by the bra $\langle 1_k|$ we get
\begin{equation}
i\hbar \frac{dc_{\!_k}(t)}{dt} = \sum_j c_{\!_j}(t) \langle 1_k|\tilde{H}(t)|1_j\rangle.
\label{ct}
\end{equation}
Using $\tilde{H}(t)$, Eq.~(\ref{ham01}), we can compute $\langle 1_k|\tilde{H}(t)|1_j\rangle$ 
via the same strategy presented in Ref. \cite{vie18}. This allows us to rewrite the 
Schr\"odinger equation as 
\begin{equation}
\frac{d\mathbf{c}(t)}{dt} = \mathbf{\tilde{M}}(t) \,\, \mathbf{c}(t),
\label{mateq}
\end{equation}
where 
\begin{equation}
\mathbf{c}(t) = \left( c_{\!_A}(t), c_{\!_1}(t), c_{\!_2}(t), \ldots, c_{\!_{N-1}}(t), 
c_{\!_N}(t), c_{\!_B}(t) \right)^T
\end{equation}
is a column vector ($T$ denotes transposition) and  
%
%%\begin{equation}
%%\mathbf{\tilde{M}}(t) =  \mathbf{M}(t) + \mathbf{P}(t)
%%\label{matriz}
%%\end{equation}
%
$\mathbf{\tilde{M}}(t)$ is the matrix of dimension $(N+2) \times (N+2)$
shown in Eq.~(\ref{matriz1}): 

\begin{equation}
{\displaystyle
\mathbf{\tilde{M}}(t) \hspace{-.01cm}=\hspace{-.01cm} -\frac{i2}{\hbar}\hspace{-.05cm}
}
\left(\hspace{-.03cm}
\begin{array}{cccccccc}
{\displaystyle D_{A}(t)} \hspace{.251cm}  & \hspace{.251cm} {\displaystyle 0} \hspace{.251cm} & \hspace{.251cm} {\displaystyle J_{\!_{A,2}}(t) } \hspace{.251cm} & \hspace{.251cm} {\displaystyle 0} \hspace{.251cm} & \hspace{.251cm} {\displaystyle 0} \hspace{.251cm} & \hspace{.251cm} {\displaystyle 0} \hspace{.251cm} & \hspace{.251cm}\cdots \\
{\displaystyle 0} \hspace{.251cm} & \hspace{.251cm} {\displaystyle D_{1}(t)} \hspace{.251cm} & \hspace{.251cm} {\displaystyle J_{\!_{1,2}}(t)} \hspace{.251cm} & \hspace{.251cm} {\displaystyle 0} \hspace{.251cm} & \hspace{.251cm} {\displaystyle 0} \hspace{.251cm} & \hspace{.251cm} {\displaystyle 0} \hspace{.251cm} & \hspace{.251cm} \cdots \\
{\displaystyle J_{\!_{A,2}}(t)} \hspace{.251cm} & \hspace{.251cm} {\displaystyle J_{\!_{1,2}}(t)} \hspace{.251cm} & \hspace{.251cm} {\displaystyle D_{2}(t)} \hspace{.251cm} & \hspace{.251cm} {\displaystyle J_{\!_{2,3}}(t)} \hspace{.251cm} & \hspace{.251cm} {\displaystyle 0} \hspace{.251cm} & \hspace{.251cm} {\displaystyle 0} \hspace{.251cm} & \hspace{.251cm} \cdots \\
{\displaystyle 0} \hspace{.251cm} & \hspace{.251cm} {\displaystyle 0} \hspace{.251cm} & \hspace{.251cm} {\displaystyle J_{\!_{2,3}}(t)} \hspace{.251cm} & \hspace{.251cm} {\displaystyle D_{3}(t)} \hspace{.251cm} & \hspace{.251cm} {\displaystyle J_{\!_{3,4}}(t)} \hspace{.251cm} & \hspace{.251cm} {\displaystyle 0} \hspace{.251cm} & \hspace{.251cm} \cdots  \\
{\displaystyle 0} \hspace{.251cm} & \hspace{.251cm} {\displaystyle 0} \hspace{.251cm} & \hspace{.251cm} {\displaystyle 0} \hspace{.251cm} & \hspace{.251cm} {\displaystyle J_{\!_{3,4}}(t)} \hspace{.251cm} & \hspace{.251cm} {\displaystyle D_{4}(t)} \hspace{.251cm} & \hspace{.251cm} {\displaystyle J_{\!_{4,5}}(t)} \hspace{.251cm} & \hspace{.251cm}\cdots  \\
{\displaystyle 0} \hspace{.251cm} & \hspace{.251cm} {\displaystyle 0} \hspace{.251cm} & \hspace{.251cm} {\displaystyle 0} \hspace{.251cm} & \hspace{.251cm} {\displaystyle 0} \hspace{.251cm} & \hspace{.251cm} {\displaystyle J_{\!_{4,5}}(t)} \hspace{.251cm} & \hspace{.251cm} {\displaystyle D_{5}(t)} \hspace{.251cm} & \hspace{.251cm}  \cdots  \\
\vdots \hspace{.251cm} & \hspace{.251cm} \vdots \hspace{.251cm} & \hspace{.251cm} \vdots \hspace{.251cm} & \hspace{.251cm} \vdots \hspace{.251cm} & \hspace{.251cm} \vdots \hspace{.251cm} & \hspace{.251cm} \vdots \hspace{.251cm}  & \hspace{.251cm}\ddots  \\
\end{array}
\hspace{-.03cm}\right)\hspace{-.015cm}.
\label{matriz1}
\end{equation}

The diagonal elements of Eq.~(\ref{matriz1}) are 
\begin{eqnarray}
D_A(t) \hspace{-0cm}&=&\hspace{-0cm} h_A(t) + \Delta(t)/2 - \Delta_{A,2}(t), \\
D_1(t) \hspace{-0cm}&=&\hspace{-0cm} h_1(t) + \Delta(t)/2 - \Delta_{1,2}(t), \\
D_2(t) \hspace{-0cm}&=&\hspace{-0cm} h_2(t) + \Delta(t)/2 - \Delta_{A,2}(t)  -\Delta_{1,2}(t)-\Delta_{2,3}(t), \\
D_j(t) \hspace{-0cm}&=&\hspace{-0cm} h_j(t) + \Delta(t)/2 - \Delta_{j-1,j}(t) 
- \Delta_{j,j+1}(t), \hspace{.1cm} \mbox{for} \hspace{.1cm} 3\leq j\leq N\!-\!2,  \\
D_{N-1}(t) \hspace{-0cm}&=&\hspace{-0cm} h_{N-1}(t)\! +\! \Delta(t)/2 \!-\! \Delta_{N-2,N-1}(t) 
\!-\! \Delta_{N-1,N}(t)\!-\!\Delta_{N-1,B}(t), \\
D_N(t) \hspace{-0cm}&=&\hspace{-0cm} h_N(t) + \Delta(t)/2 - \Delta_{N-1,N}(t), \\
D_B(t) \hspace{-0cm}&=&\hspace{-0cm} h_B(t) + \Delta(t)/2 - \Delta_{N-1,B}(t), 
\end{eqnarray}
with
\begin{equation}
\Delta(t) = \Delta_{A,2}(t)+\sum_{j=1}^{N-1}\Delta_{j,j+1}(t) + \Delta_{N-1,B}(t).
\end{equation}

The solution to Eq.~(\ref{mateq}) can be formally written as
a time-ordered matrix exponential with the aid of the time-ordering operator 
$\hat{\mathcal{T}}$,
\begin{equation}
\mathbf{c}(t) = \hat{\mathcal{T}}\exp\left(\int_0^t\mathbf{\tilde{M}}(t')dt'\right)\mathbf{c}(0),
\label{formalsolution}
\end{equation}
where $\mathbf{c}(0)$ is the column vector containing the initial conditions listed 
in Eq.~(\ref{initial1}).

\section{Quantifying entanglement}
\label{eof}

We use the Entanglement of Formation (EoF) \cite{ben96,woo98} to quantify 
the entanglement shared between two qubits. Specifically, we want to determine 
as a function of time the EoF between qubits $N$ and $B$ with Bob. Tracing 
out the other $N$ qubits from $\rho(t)=|\Psi(t)\rangle \langle\Psi(t)|$, 
the density matrix describing the whole system, we get 
\cite{vie18}
\begin{eqnarray}
\hspace{-.15cm}\rho_{\!_{NB}}(t) 
&=& 
\left(\hspace{-.0cm}
\begin{array}{cccc}
1-|c_{\!_N}(t)|^2-|c_{\!_B}(t)|^2 \hspace{-.125cm}&\hspace{-.125cm} 0 \hspace{-.125cm}&\hspace{-.125cm} 0 & 0 \\
0 \hspace{-.125cm}&\hspace{-.125cm} |c_{\!_B}(t)|^2 \hspace{-.125cm}&\hspace{-.125cm} c_{\!_B}(t)c_{\!_N}^*(t) & 0 \\
0 \hspace{-.125cm}&\hspace{-.125cm} c_{\!_N}(t)c_{\!_B}^*(t) \hspace{-.125cm}&\hspace{-.125cm} |c_{\!_N}(t)|^2 & 0 \\
0 \hspace{-.125cm}&\hspace{-.125cm} 0 \hspace{-.125cm}&\hspace{-.125cm} 0 & 0
\end{array}
\hspace{-.0cm}
\right), \nonumber \\
\label{rhoNB}
\end{eqnarray}
where $*$ means complex conjugation. Eq. (\ref{rhoNB}) is the density matrix describing the two qubits 
$N$ and $B$ with Bob written in the basis $\{|00\rangle,|01\rangle,|10\rangle,|11\rangle \}$. 

With the aid of $\rho_{\!_{NB}}(t)$ the EoF can be computed following the recipe in \cite{woo98,vie18},
\begin{equation}
\mbox{EoF}(\rho_{\!_{NB}}) = -f(C)\log_2f(C) - [1-f(C)]\log_2[1-f(C)], 
\label{eof2}
\end{equation}
with $f(C)=(1 + \sqrt{1 - C^2})/2$ and concurrence $C$ given by  
$
C(t) = 2|c_{\!_N}(t)c_{\!_B}(t)|.
$
Note that a maximally entangled two-qubit state has EoF = 1 and 
when the pair of qubits shares no entanglement EoF = 0.

\section{Modeling disorder and noise}
\label{results}

We deal with the three major types of disorder that can affect the system's coupling
constants $J_{i,j}$, namely, static, dynamic, and fluctuating disorders \cite{vie13,vie14}, and assume 
that any fluctuation in those coupling constants is about the optimal settings
for the ordered case with $N=100 +2$ qubits.  Specifically, we either employ $J_m/J=2.86$, the optimal setting
for $J_m/J\leq 5.0$ and whose transmitted entanglement is 
EoF $= 0.816$, or $J_m/J=49.98$, the best case when $J_m/J\leq 50.0$ and with transmitted
entanglement given by EoF $= 0.986$ \cite{vie18} (see the Appendix for calculations related to chains of $N=1000 + 2$
qubits).
In the rest of this work we set $J=1.0$ and $\hbar = 1.0$, which
defines $J$ as our unit of energy, and work with the two values of $J_m$ 
given above.
Following Ref. \cite{vie13,vie14}, we also assume that when dealing with time dependent disorder,
the Hamiltonian changes with time only a finite number of times and after the same period $\tau$.

In this scenario, we can model disorder if we set 
\begin{equation}
J_{i,j}(t_k) = J_{i,j}(t_{k-1})\left[ 1 + \delta J_{i,j}(k) \right],
\label{change1}
\end{equation}
with
$J_{A,2}(0) = J_{1,2}(0) = J_{N-1,N}(0) = J_{N-1,B}(0) = J,$ 
$\sum_{j=2}^{N-2}J_{j,j+1}(0) = J_m,$ and  
$J_{i,j}(t) = 0$, any $t$, for other values of $i,j$.
%
%%\begin{eqnarray}
%%J_{A,2}(0) = J_{1,2}(0) = J_{N-1,N}(0) = J_{N-1,B}(0) = J,  \\
%%\sum_{j=2}^{N-2}J_{j,j+1}(0) = J_m, \\
%%J_{i,j}(t) = 0, \mbox{any $t$, for other values of}\; i,j.
%%\end{eqnarray}
%
Here 
\begin{equation}
(k-1)\tau < t_k \leq k\tau,      
\end{equation}
where $k$ is an integer such that $1\leq k \leq n$.
We also have $t_0=0$ and we define $T_{max}=\max\{t_n\}=n\tau$. Sometimes we call $T_{max}$ by $Jt/\hbar$, both notations meaning
the time when in the optimal ordered model the transmitted entanglement from Alice 
to Bob is maximal.
The quantity $\delta J_{i,j}(k)$ is responsible for introducing
disorder in our system. Its behavior depends on which type of disorder we have 
but all cases boil down to random continuous uniform distributions 
centered in zero and whose ranges lie between $-p$ and $p$, with 
$p$ the maximal percentage fluctuation 
of $J_{i,j}$ about its optimal value. Note that $\delta J_{i,j}(k)$
is a function of the index $k$, i.e., whenever $k$ changes we have a new uniform 
distribution from which the random values of $\delta J_{i,j}$ are drawn. 

In a similar way we can introduce noise. 
Noting that the optimal settings have no external magnetic field ($h_j(t)=0$) and
no $\sigma_i^z\sigma_j^z$ interaction ($\Delta_{i,j}(t)=0$), the following
prescription allows us to insert noise in our system,
\begin{eqnarray}
h_j(t_k) &=& h_j(t_{k-1}) + \delta h_j(k), \label{change2}\\ 
\Delta_{i,j}(t_k) &=& \Delta_{i,j}(t_{k-1}) + \delta \Delta_{i,j}(k),
\label{change3}
\end{eqnarray}
where $t_k$ and $k$ were defined above, $h_j(0)=0$ for $j=A,1,$ $2,\ldots, N-1,N,B$, $\Delta_{A,2}(0)$ $=$ $\Delta_{1,2}(0)=\Delta_{N-1,N}(0)=\Delta_{N-1,B}(0)=0$,
$\sum_{j=2}^{N-2}\Delta_{j,j+1}(0)=0$, and $\Delta_{i,j}(t)=0$, any $t$, for other values
of $i,j$.
Analogously to $\delta J_{i,j}(k)$, $\delta h_j(k)$ or $\delta \Delta_{i,j}(k)$ 
depends on the kind of disorder we have and will be explained shortly. But 
as before these uniform distributions are centered in zero, ranging  
between $-p$ and $p$, where $p$ now means the greatest possible value of  
$\delta h_j(k)$ and $\delta \Delta_{i,j}(k)$. Since our unit of energy is $J=1.0$,
$p$ can be read as the percentage of $J$ 
defining the upper bound for 
$\delta h_j(k)$ or $\delta \Delta_{i,j}(k)$ in a given uniform distribution.

\subsection{Static disorder} 

In this scenario we have noise or fluctuations in
the coupling constants that are time independent but site (position) dependent,
\begin{eqnarray}
&&\hspace{-.75cm} \delta J_{i,j}(1) = \delta J_{i,j},  \hspace{.25cm} \delta h_j(1) = \delta h_j, \hspace{.25cm} 
\delta\Delta_{i,j}(1) = \delta \Delta_{i,j}, \\
&&\delta J_{i,j}(k)=\delta h_j(k)=\delta\Delta_{i,j}(k)=0, \hspace{.5cm} k\neq 1.
\label{static}
\end{eqnarray}
This means that 
$J_{i,j}(t) =  J_{i,j}(0)\left( 1 + \delta J_{i,j}\right)$, $h_j(t) = \delta h_j$, 
$\Delta_{i,j}(t) = \delta \Delta_{i,j}$, where $0\leq t \leq n\tau$. 
In other words, each quantity above is chosen from independent continuous uniform
distributions at $t = 0$ and fixed during the time evolution. 

In this case 
Eq.~(\ref{formalsolution}) can be integrated and we get
\begin{equation}
\mathbf{c}(t) = \exp(\mathbf{\tilde{M}}\,\, t )\mathbf{c}(0).
\label{formalsolution2}
\end{equation}
By numerically computing the matrix exponential we obtain $\mathbf{c}(t)$
and then, using Eq.~(\ref{eof2}), the entanglement at any $t$ 
between the qubits $N$ and $B$ with Bob.

\subsection{Dynamic disorder} 

In contrast to the case of static disorder, we have time dependent but position independent noise
or fluctuations in the coupling constants,
\begin{eqnarray}
\hspace{-.5cm}\delta J_{i,j}(k) = \delta J(k), &\hspace{-.0cm} \delta h_j(k) = \delta h(k), &\hspace{-.0cm} \delta \Delta_{i,j}(k) = \delta \Delta(k).
\label{dynamic}
\end{eqnarray}
Now, each coupling constant changes equally by $\delta J(k)$ at the time $(k-1)\tau$, remaining fixed from $(k-1)\tau$ to
$k\tau$. After each period of time $\tau$ a new random number is drawn from a continuous uniform distribution as previously explained
and a new value of the coupling constant is obtained: $J_{i,j}(t_k) = J_{i,j}(t_{k-1})\left[ 1 + \delta J(k) \right]$. 
Similarly, all qubits are subjected to the
same external field from $(k-1)\tau$ to $k\tau$ and the $\sigma_i^z\sigma_j^z$ interaction strength 
is the same for any
pair of nearest neighbors during this time span. After a period of time $\tau$ new values for the field and for the 
$\sigma_i^z\sigma_j^z$ interaction strengths are obtained from two independent continuous uniform distributions
$\delta h(k)$ and $\delta \Delta(k)$ according to the prescription given in Eq.~(\ref{change2}) and (\ref{change3}).
We should note that it is convenient sometimes to express $\tau$ as a percentage
of $T_{max}=Jt/\hbar$. Thus $\tau = (Jt/\hbar)/n$, with $1/n$ denoting the percentage of the total time of evolution
$T_{max}$ giving the period of changes. 

Noting that between time $(k-1)\tau$ and $k\tau$, where $k$ is an integer 
between $1$ and $n$, the matrix $\mathbf{\tilde{M}}(t)$ is time independent, the time evolution
of the system all the way down to $T_{max}=n\tau$ can be written as
\begin{equation}
\mathbf{c}(n\tau) = \exp(\mathbf{\tilde{M}_n}\,\, \tau )\exp(\mathbf{\tilde{M}_{n-1}}\,\, \tau )
\ldots\exp(\mathbf{\tilde{M}_1}\,\, \tau )\mathbf{c}(0), 
\label{formalsolution3}
\end{equation}
where $\mathbf{\tilde{M}_k}$ is a time independent matrix given by Eq.~(\ref{matriz1})
such that $\mathbf{\tilde{M}_k} = \mathbf{\tilde{M}_{k-1}} + \delta\mathbf{\tilde{M}_k}$.
Here $\mathbf{\tilde{M}_{0}}$ is associated to the optimal ordered Hamiltonian, 
Eq.~(\ref{ham0}), and $\delta\mathbf{\tilde{M}_k}$ is the modification to $\mathbf{\tilde{M}_{k-1}}$
coming from the presence of the dynamical disorder given by Eqs.~(\ref{change1})-(\ref{change3}) 
and (\ref{dynamic}).

\subsection{Fluctuating disorder} 

This is the case where both aspects of static and dynamic disorders are combined, namely,
where we have time and position dependent noise or time and position 
dependent fluctuations in the coupling constants. The coupling constants between qubits as well as 
the strength of the noise are given by Eqs.~(\ref{change1})-(\ref{change3}), with
$\delta J_{i,j}(k), \delta h_j(k),$ and $\delta \Delta_{i,j}(k)$ depending on the position of the qubits
and on time in the same way already described, respectively, for the cases of static and dynamic disorders.
The time evolution of the system is given by Eq.~(\ref{formalsolution3}) 
and $\mathbf{\tilde{M}_k}$ is a time independent matrix with  
position dependent coefficients given by Eq.~(\ref{matriz1}).

\section{Results}
\label{robustness}

\begin{figure}[!ht]
\begin{center}
\includegraphics[width=11.8cm]{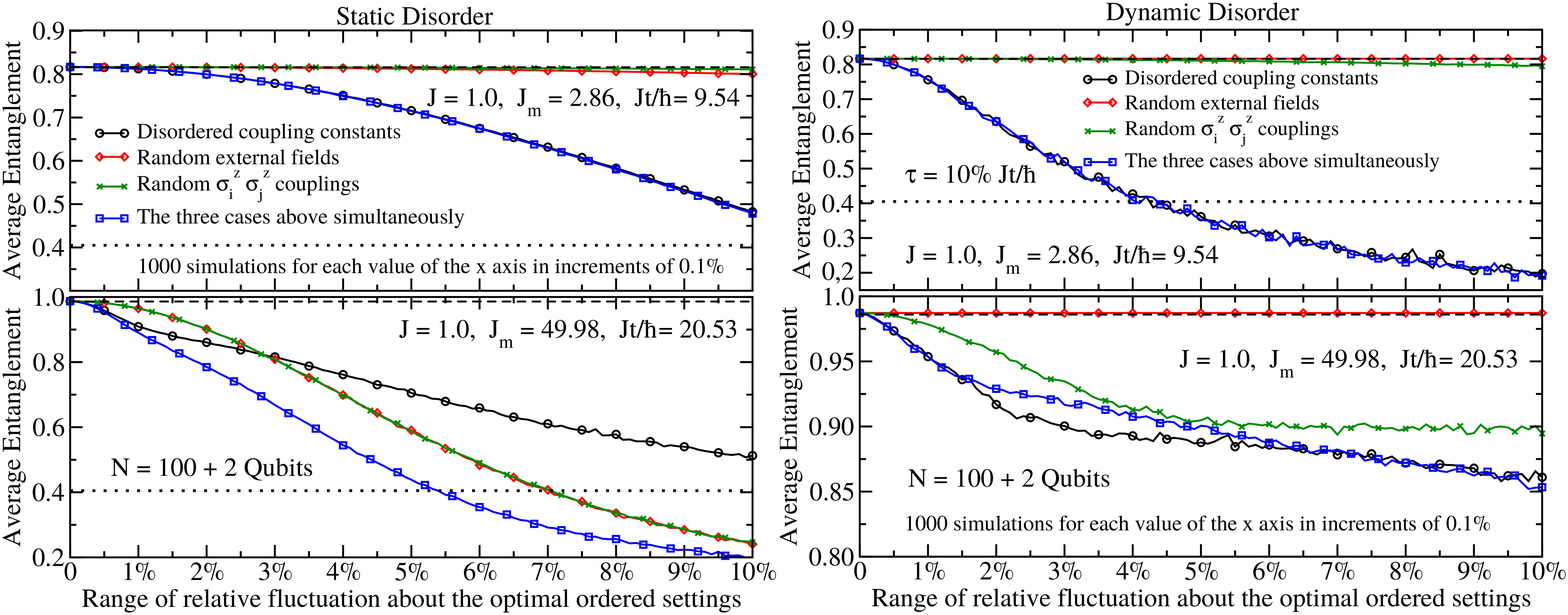}
\caption{
%\label{fig2e3e4} 
\label{fig2e3}
Proposed model with $N= 100 + 2$ qubits with 
$1000$ static (left) and dynamic (right) %and fluctuating (right) 
disorder realizations for each value of the x-axis (percentage deviation $p$ as explained in the text), 
starting at $0.1\%$ and going up to $p=10\%$, in 
increments of $0.1\%$. 
All solid curves and the dashed ones refer to the model here reported (proposed model). The only curves
referring to the standard model (strictly one dimensional chain) are the dotted ones.
Upper panels: The solid curves
give the average of the entanglement transmitted
after $1000$ realizations for each value of $p$ and at $Jt/\hbar = 9.54$, 
the time where the optimal transmission of entanglement in the ordered system 
occurs if $J_m/J = 2.86$. This value for $J_m/J$ is that giving the best performance
when $J_m/J$ ranges from $0$ to $5$.
The circle-black curves denote the noiseless case, where only the coupling constants are subjected to 
disorder. The diamond-red curves represent the case where the noise is given by 
external magnetic fields and the star-green ones when noise is given by the $\sigma_i^z\sigma_j^z$
interaction. The square-blue curves depict the case where both noises are present as well as
disorder in the coupling constants.
The dashed curves are the optimal entanglement 
transmitted for the clean system and the dotted ones show the greatest entanglement 
transmitted for the standard $N=100$ qubits ordered model.
Lower panels: Again we have the proposed model but now $Jt/\hbar = 20.53$, 
the time where the optimal transmission of entanglement for the  
clean system occurs when $J_m/J = 49.98$. This value for $J_m/J$ is the one giving the best performance
for the proposed model when $J_m/J$ ranges from $0$ to $50$. See text for more details.
}
\end{center} 
\end{figure}

\begin{figure}[!ht]
\begin{center}
\includegraphics[width=8.5cm]{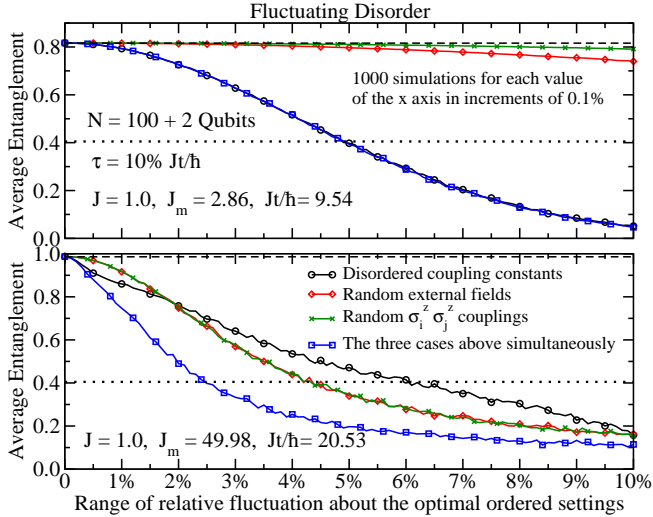}
\caption{
\label{fig4_alt} 
All quantities similarly defined as explained in the caption of Fig. \ref{fig2e3}, with the proviso that
we now have fluctuating disorder. Note that all curves refer to the proposed model, with the exception of 
the dotted ones, which are the optimal entanglement transmitted for the ordered standard model
(strictly one dimensional chain).
Upper panel: The entanglement reaching Bob is measured at the time $Jt/\hbar = 9.54$, 
the time where the optimal transmission of entanglement for the  
clean proposed model occurs when $J_m/J = 2.86$. This value for $J_m/J$ is the one giving the best performance
for the ordered proposed model when $J_m/J$ ranges from $0$ to $5$ and was the value we attributed to the coupling constants
of the Hamiltonian at $t=0$.
Lower panel: The entanglement reaching Bob is measured at the time $Jt/\hbar = 20.53$, 
the time where the optimal transmission of entanglement for the  
clean proposed model occurs when $J_m/J = 49.98$. This value for $J_m/J$ is the one giving the best performance
for the ordered proposed model when $J_m/J$ ranges from $0$ to $50$ and was the value we attributed to the coupling constants
of the Hamiltonian at $t=0$.}
\end{center} 
\end{figure}

We first study static disorder and dynamic and fluctuating disorders for
$\tau = 10\% T_{max}=10\%Jt/\hbar$ %(Fig. \ref{fig2e3e4}). 
(Figs. \ref{fig2e3} and \ref{fig4_alt}).

%\begin{figure*}[!ht] 
%\end{figure*}

This value of $\tau$ means that the Hamiltonian changes $10$ times during
the time evolution of the system from $t=0$ to $t=T_{max}$ according to the prescription 
given above. 

Looking at the upper panels of %Fig. \ref{fig2e3e4}, 
Figs. \ref{fig2e3} and \ref{fig4_alt},
we note the following trends for the entanglement 
transmitted to Bob. For small values of $J_m/J$ the presence of noise barely affects the efficiency of the protocol, 
i.e., the presence of random  external magnetic fields and $\sigma_i^z\sigma_j^z$ interactions do not affect 
considerably the system's performance up to a deviation of $10\%$ about the optimal settings for the ordered case. 
On the other hand, the presence of disorder in the coupling constants affects the system performance much more strongly than the 
presence of noise. However, for deviations of $1\%$ about the optimal settings, the efficiency is barely reduced.

Moreover, among the three types of disorder in the coupling constants, static disorder is the least severe. 
In this scenario, even for a deviation of 
$10\%$ about the optimal settings, we still beat the entanglement transmitted using the optimal ordered standard model (dotted lines).
Note that dynamic and fluctuating disorders in the coupling constants affect the system nearly in the same way. 
When noise and disorder in the coupling constants act simultaneously, the reduction in efficiency is almost the same as if the
noise was absent, another confirmation of the fact that the types of noise here studied barely affect the system's efficiency
for small values of $J_m/J$.

\begin{figure}[!ht] 
\begin{center}
\includegraphics[width=8.5cm]{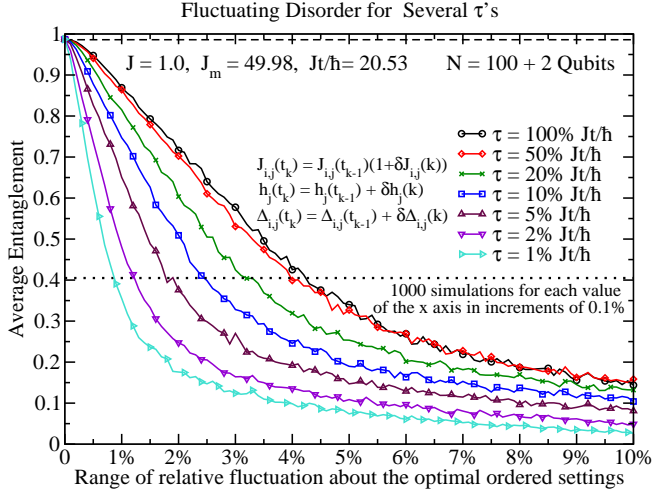}
\caption{
\label{fig5} 
Proposed model with noise and coupling constants simultaneously modeled
by fluctuating disorder. Here $J_m/J = 49.98$ and we study how the efficiency of
the entanglement transmission is
affected as we decrease $\tau$, the period of changes in the Hamiltonian.
The meaning of all curves above and how they were computed are
similar to those given in 
%Fig. \ref{fig2e3e4}. 
Figs. \ref{fig2e3} and \ref{fig4_alt}.
It is clear that the smaller the period of changes, the lower the amount of entanglement
reaching Bob. Note that all curves refer to the proposed model, with the exception of 
the dotted one, which is the optimal entanglement transmitted for the ordered standard model
(strictly one dimensional chain).}
\end{center} 
\end{figure}

Moving to the lower panels of %Fig. \ref{fig2e3e4}, 
Figs. \ref{fig2e3} and \ref{fig4_alt},
we note different trends for the entanglement 
transmitted to Bob. 
Indeed, for high values of $J_m/J$ the presence of noise becomes relevant in almost all cases, reducing the efficiency of the protocol,
and the presence of disorder in the coupling constants no longer dominates the reduction of the system's performance. 
Of all three types of disorder, dynamic disorder is now the least severe. In this case, even for a deviation of 
$10\%$ about the optimal settings, the entanglement transmitted to Bob is very high, close to the optimal value for the 
clean system and way above the entanglement reaching Bob if we employ the standard model (dotted lines).
Dynamic and fluctuating disorders, on the other hand, no longer affect the system in the same way,
with the latter being much more severe.
When noise and disorder in the coupling constants act simultaneously, we get almost always the worst scenario, 
where the reduction in efficiency is most severe (square-blue curves). 
However, even for high values of $J_m/J$, deviations of $1\%$ about the optimal settings do
not affect much the amount of entanglement reaching Bob.

In Fig. \ref{fig5} we study how the rate of change of the Hamiltonian during the time evolution affects the system's 
performance. We work with the worst of the six scenarios shown in %Fig. \ref{fig2e3e4} (lower-right panel), 
Figs. \ref{fig2e3} and \ref{fig4_alt},
namely, fluctuating
disorder with noise and disorder in the coupling constants simultaneously affecting the system. 
Looking at Fig. \ref{fig5} it is not difficult to convince ourselves that the higher the frequency of changes in the
Hamiltonian or, equivalently, the smaller the period $\tau$ of changes,  
the lower the entanglement transmitted to Bob. 

The reason why a small period in the fluctuating disorder leads to a poor transmission of entanglement 
is the fact that the smaller the period of changes, the higher the frequency of changes in the Hamiltonian.
This higher number of changes in the Hamiltonian is cumulative and ultimately leads to a greater deviation from 
the optimal values of the coupling constants. 
For longer periods, nevertheless, the Hamiltonian changes only once or twice
during the whole time evolution, leading to almost no change in the optimal settings when we work with small disorder. 
This fact reflects in a better performance for systems with longer periods of changes in the Hamiltonian when
compared with the shorter ones.

\begin{figure}[!ht] 
\begin{center}
\includegraphics[width=8.5cm]{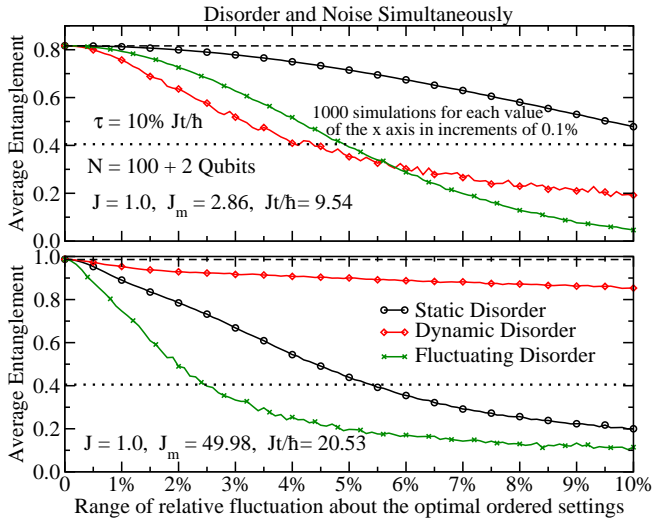}
\caption{
\label{fig6} 
We show for two values of $J_m/J$ how different types of disorder affect the proposed model's performance.
The meaning of all quantities shown here is the same as explained
in Figs. %\ref{fig2e3e4} 
\ref{fig2e3}, \ref{fig4_alt}, and \ref{fig5}.
Note that all curves refer to the proposed model, with the exception of 
the dotted ones, which are the optimal entanglement transmitted for the ordered standard model
(strictly one dimensional chain).
}
\end{center} 
\end{figure}

In order to make it clearer to analyze how different types of disorder affect the efficiency of the protocol, 
we show in Fig. \ref{fig6} and in the same graphic the entanglement transmitted for the 
three different types of disorder studied here. We fix 
the value of $\tau$ such that it is $10\%$ of $Jt/\hbar$, when dealing with dynamic and fluctuating disorders,
and work with the scenario where noise and disorder in the coupling constants are simultaneously affecting the system.

For $J_m/J = 2.86$, upper panel of Fig. \ref{fig6}, we note that for small fluctuations about the optimal settings
($<5\%$), dynamic disorder is the most severe type of disorder. For greater values of deviations about the optimal
settings, fluctuating disorder becomes the worst case. Also, for small values of $J_m/J$, static disorder is the 
least severe type of disorder.
Looking at the lower panel of Fig. \ref{fig6}, when $J_m/J = 49.98$, we note that dynamic disorder becomes very
benign, little affecting the system's performance. We also note that fluctuating disorder is the most severe 
type of disorder for all range of fluctuations about the optimal settings of the clean system.

\section{Conclusion}
\label{conclusion}

Summing up, we have extensively studied how the protocol presented in Ref. \cite{vie18},
specially devised to transmit maximally entangled Bell states, responds to disorder and 
noise for systems of size of the order of a hundred qubits. 
We have studied three types of disorder, namely, static, dynamic, and fluctuating disorders 
\cite{vie13,vie14} and two types of noise that may affect the system. In one type of noise we have 
random external magnetic fields acting on the qubits of the system and in the 
other one we investigated how spurious $\sigma^z_i\sigma^z_j$ interactions reduce the system's
capacity to transmit entanglement. 

For all the several cases of noise and disorder 
presented here, we showed that for deviations of up to $1\%$ about the optimal settings of the clean
system, we still have excellent entanglement transmission, with little or no reduction in the efficiency
of the protocol. Moreover, up to deviations about the
optimal settings of the order of  $3\%$, the proposed model (upper panel of Fig. \ref{fig1})
still beats the optimal entanglement transmission efficiency of the 
standard model (lower panel of Fig. \ref{fig1}). For a system's size of the order of
a thousand qubits, we obtain excellent transmission of entanglement for deviations in the optimal
settings of up to $0.1\%$ and acceptable ones for deviations of up to $0.5\%$ (see the Appendix).   

All those results show that the proposed model is very robust to
noise and disorder and we believe it might prove fruitful to test in future works its robustness when
the number of excitations is no longer conserved and when we have residual Dzyaloshinskii-Moriya interactions
affecting the system \cite{dzy58,mor60}.

%\begin{acknowledgments}
%RV thanks CNPq (Brazilian National Council for Scientific and Technological Development)
%for funding and GR thanks CNPq and CNPq/FAPERJ (State of Rio de Janeiro Research Foundation) for financial support through the National Institute of
%Science and Technology for Quantum Information.
%\end{acknowledgments}

\section*{Acknowledgments}
RV thanks CNPq (Brazilian National Council for Scientific and Technological Development)
for funding and GR thanks CNPq and CNPq/FAPERJ (State of Rio de Janeiro Research Foundation) for financial support through the National Institute of
Science and Technology for Quantum Information.

\appendix

\section{The 1000 + 2 qubits case}
\label{Appendix}

Whenever we have dynamic or fluctuating disorders, specially for small values of $\tau$, the computational resources 
needed to handle systems of the order of thousands of qubits are very demanding. 
For that reason, we report here only two calculations for the $N = 1000 + 2$ qubits system by implementing  
one tenth of the realizations made for the $N = 100 + 2$ qubits case, namely, instead of $1000$ realizations for
each disorder percentage, we work with $100$ realizations. We have also worked with the 
worst possible scenario, in which the system is most severely affected, i.e., fluctuating disorder affecting the coupling constants as well
as fluctuating random external fields and spurious $\sigma^z_i\sigma^z_j$ interactions. Note that for less stringent
disorder and noise scenarios, the $N = 1000 + 2$ qubits case gives much better results.

\begin{figure}[!ht] 
\begin{center}
\includegraphics[width=8.5cm]{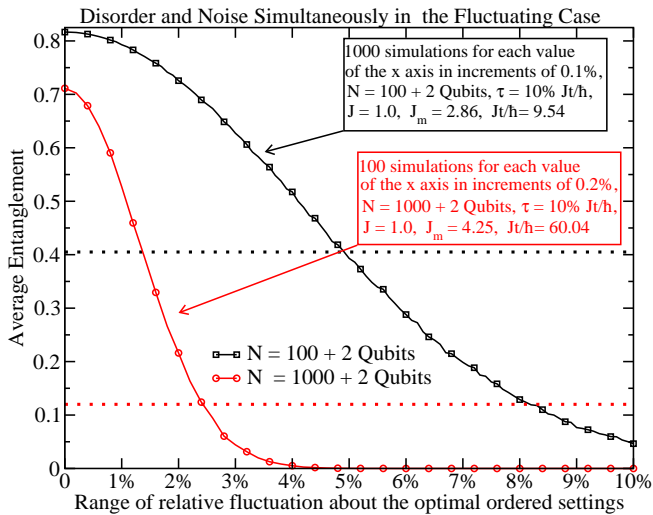}
\caption{
\label{fig7} 
For $N = 100 + 2$ qubits we have $J_m/J=2.86$, the optimal setting
for the clean system when $J_m/J\leq 5.0$, giving a transmitted entanglement of 
EoF $= 0.81$ at the time $Jt/\hbar = 9.54$ (square-black curve). 
For $N = 1000 + 2$ qubits we have $J_m/J=4.25$, the optimal setting
for the clean system when $J_m/J\leq 5.0$, with transmitted entanglement given by 
EoF $= 0.71$ at the time $Jt/\hbar = 9.54$ (circle-red curve).
Solid curves are related to the proposed model when affected by fluctuating disorder and noise 
about those optimal settings while the dotted curves refer to the optimal 
entanglement transmitted if we employ the clean standard model (strictly linear chain).
The period $\tau$ of changes in the Hamiltonian is shown in the figure.
}
\end{center} 
\end{figure}

\begin{figure}[!ht] 
\begin{center}
\includegraphics[width=8.5cm]{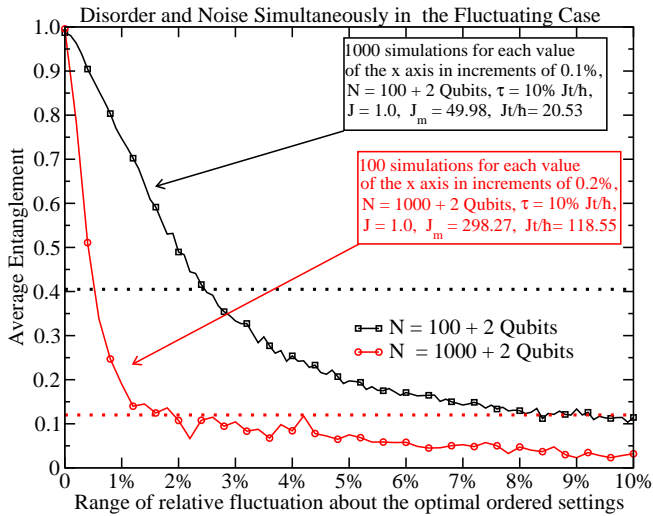}
\caption{
\label{fig8} 
For $N = 100 + 2$ qubits we have $J_m/J=49.98$, the optimal setting
for the clean system when $J_m/J\leq 50$, giving a transmitted entanglement of 
EoF $= 0.98$ at the time $Jt/\hbar = 20.53$ (square-black curve). 
For $N = 1000 + 2$ qubits we have $J_m/J=298.27$, the optimal setting
for the clean system when $J_m/J\leq 300$, with transmitted entanglement given by 
EoF $= 0.99$ at the time $Jt/\hbar = 118.55$ (circle-red curve).
Solid curves are related to the proposed model when affected by fluctuating disorder and noise 
about those optimal settings while the dotted curves refer to the optimal 
entanglement transmitted if we employ the clean standard model (strictly linear chain).
The period $\tau$ of changes in the Hamiltonian is shown in the figure.
}
\end{center} 
\end{figure}

As expected, the results depicted in Figs. \ref{fig7} and \ref{fig8} show 
that the longer the chain the more the system is affected by disorder and noise. 
However, we still have very good entanglement transmission for disorder of less than $0.1\%$ about the optimal settings
of the clean system and an acceptable one for less than $0.5\%$. Moreover, for disorder of less than $1\%$, we still
beat the optimal transmission of the clean standard model (strictly linear chain).

\pagebreak


\begin{thebibliography}{200}

\bibitem{ben00} Bennett, C. H., DiVincenzo, D. P.: Quantum information and computation. 
Nature (London) \textbf{404}, 247 (2000)

\bibitem{ben93} Bennett, C. H., Brassard, G., Cr\'epeau, C., Jozsa, R., Peres, A., Wootters, W. K.: Teleporting an unknown quantum state via dual classical and 
Einstein-Podolsky-Rosen channels. 
Phys. Rev. Lett. \textbf{70}, 1895 (1993) 

\bibitem{bos03} Bose, S.: Quantum Communication through an Unmodulated Spin Chain. 
Phys. Rev. Lett. \textbf{91}, 207901 (2003)

\bibitem{chr04} Christandl, M., Datta, N., Ekert, A., Landahl, A. J.: Perfect State Transfer in Quantum Spin Networks. 
Phys. Rev. Lett. \textbf{92}, 187902 (2004)

\bibitem{nik04} Nikolopoulos, G. M., Petrosyan, D., Lambropoulos, P. L.: Electron wavepacket propagation in a chain of coupled quantum dots. 
J. Phys.: Condens. Matter \textbf{16}, 4991 (2004)

\bibitem{sub04} Subrahmanyam, V.: Entanglement dynamics and quantum-state transport in spin chains. 
Phys. Rev. A \textbf{69}, 034304 (2004)

\bibitem{osb04} Osborne, T. J., Linden, N.: Propagation of quantum information through a spin system. 
Phys. Rev. A \textbf{69}, 052315 (2004)

\bibitem{cir97} Cirac, J. I., Zoller, P., Kimble, H., Mabuchi, H.: Quantum State Transfer and Entanglement Distribution 
among Distant Nodes in a Quantum Network. 
Phys. Rev. Lett. \textbf{78}, 3221 (1997)

\bibitem{ple04} Plenio, M. B., Hartley, J., Eisert, J.: Dynamics and manipulation of entanglement in coupled harmonic systems with many degrees of freedom. 
New J. Phys. \textbf{6}, 36 (2004)

\bibitem{sem05} Plenio, M. B., Semi\~ao, F. L.: High efficiency transfer of quantum information and multiparticle entanglement generation 
in translation-invariant quantum chains. New J. Phys. \textbf{7}, 73 (2005)

\bibitem{chr05} Christandl, M., Datta, N., Dorlas, T. C., Ekert, A., Kay, A., Landahl, A. J.: Perfect transfer of arbitrary states in quantum spin networks. 
Phys. Rev. A \textbf{71}, 032312 (2005)

\bibitem{shi05} Shi, T., Li, Y., Song, Z., Sun, Ch.-P.: Quantum-state transfer via the ferromagnetic chain in a spatially modulated field. 
Phys. Rev. A \textbf{71}, 032309 (2005)

\bibitem{woj05} W\'ojcik, A., \L{}uczak, T., Kurzy\'nski, P., Grudka, A., Gdala, T., Bednarska, M.: Unmodulated spin chains as universal quantum wires. 
Phys. Rev. A \textbf{72}, 034303 (2005)

\bibitem{li05} Li, Y., Shi, T., Chen, B., Song, Z., Sun, C.-P.: Quantum-state transmission via a spin ladder as a robust data bus. 
Phys. Rev. A \textbf{71}, 022301 (2005)

\bibitem{kar05} Karbach, P., Stolze, J.: Spin chains as perfect quantum state mirrors. 
Phys. Rev. A \textbf{72}, 030301 (2005)

\bibitem{kay05} Kay, A., Ericsson, M.: Geometric effects and computation in spin networks. 
New J. Phys. \textbf{7}, 143 (2005)

\bibitem{har06} Hartmann, M. J., Reuter, M. E., Plenio, M. B.: Excitation and entanglement transfer versus spectral gap. 
New J. Phys. \textbf{8}, 94 (2006)

\bibitem{huo08} Huo, M. X., Li, Y., Song, Z., Sun, C. P.: The Peierls distorted chain as a quantum data bus for quantum state transfer. 
Europhys. Lett. \textbf{84}, 30004 (2008)

\bibitem{gua08} Gualdi, G., Kostak, V., Marzoli, I., Tombesi, P.: Perfect state transfer in long-range interacting spin chains. 
Phys. Rev. A \textbf{78}, 022325 (2008)

\bibitem{ban10} Banchi, L., Apollaro, T. J. G., Cuccoli, A., Vaia, R., Verrucchi, P.: Optimal dynamics for quantum-state and entanglement 
transfer through homogeneous quantum systems. 
Phys. Rev. A \textbf{82}, 052321 (2010)

\bibitem{kur11} Kurzy\'nski, P., W\'ojcik, A.: Discrete-time quantum walk approach to state transfer. 
Phys. Rev. A \textbf{83}, 062315 (2011)

\bibitem{kay11} Pemberton-Ross, P. J., Kay, A.: Perfect Quantum Routing in Regular Spin Networks. 
Phys. Rev. Lett. \textbf{106}, 020503 (2011)
  
\bibitem{god12} Godsil, C., Kirkland, S., Severini, S., Smith, J.: Number-Theoretic Nature of Communication in Quantum Spin Systems. 
Phys. Rev. Lett. \textbf{109}, 050502 (2012)

\bibitem{apo12} Apollaro, T. J. G., Banchi, L., Cuccoli, A., Vaia, R., Verrucchi, P.: 99\%-fidelity ballistic quantum-state transfer 
through long uniform channels. 
Phys. Rev. A \textbf{85}, 052319 (2012)

\bibitem{lor13} Lorenzo, S., Apollaro, T. J. G., Sindona, A., Plastina, F.: Quantum-state transfer via resonant tunneling through local-field-induced barriers.
Phys. Rev. A \textbf{87}, 042313 (2013)

\bibitem{sou14} Sousa, R., Omar, Y.: Pretty good state transfer of entangled states through quantum spin chains. 
New J. Phys. \textbf{16}, 123003 (2014)

\bibitem{hor14} Korzekwa, K., Machnikowski, P., Horodecki, P.: Quantum-state transfer in spin chains via isolated resonance of terminal spins. 
Phys. Rev. A \textbf{89}, 062301 (2014)

\bibitem{shi15} Shi, Z. C., Zhao, X. L., Yi, X. X.: Robust state transfer with high fidelity in spin-1/2 chains by Lyapunov control. 
Phys. Rev. A \textbf{91}, 032301 (2015)

\bibitem{lor15} Lorenzo, S., Apollaro, T. J. G., Paganelli, S., Palma, G. M., Plastina, F.: Transfer of arbitrary two-qubit states via a spin chain. 
Phys. Rev. A \textbf{91}, 042321 (2015)

\bibitem{pou15} Pouyandeh, S., Shahbazi, F.: Quantum state transfer in XXZ spin chains: A measurement induced transport method. 
Int. J. Quantum. Inform. \textbf{13}, 1550030 (2015)

\bibitem{zha16} Zhang, X.-P., Shao, B., Hu, S., Zou, J., Wu, L.-A.: Optimal control of fast and high-fidelity quantum state transfer in spin-1/2 chains. 
Ann. Phys. (NY) \textbf{375}, 435 (2016)

\bibitem{che16} Chen, X., Mereau, R., Feder, D. L.: Asymptotically perfect efficient quantum state transfer across uniform chains with two impurities. 
Phys. Rev. A \textbf{93}, 012343 (2016)

\bibitem{nic16} Nicacio, F., Semi\~ao, F. L.: Transport of correlations in a harmonic chain. Phys. Rev. A \textbf{94}, 012327 (2016)

\bibitem{est17}  Estarellas, M. P., D'Amico, I., Spiller, T. P.: Topologically protected localised states in spin chains. Sci. Rep. \textbf{7}, 42904 (2017)
 
\bibitem{est17b} Estarellas, M. P., D'Amico, I., Spiller, T. P.: Robust quantum entanglement generation and generation-plus-storage protocols with spin chains.
Phys. Rev. A \textbf{95}, 042335 (2017)

\bibitem{apo18} Apollaro, T. J. G., Almeida, G. M. A., Lorenzo, S., Ferraro, A., Paganelli, S.: Spin chains for two-qubit teleportation.
Available from: arXiv:1812.11609 [quant-ph] (2018).
 
\bibitem{chi05} De Chiara, G., Rossini, D., Montangero, S., Fazio, R.: From perfect to fractal transmission in spin chains. 
Phys. Rev. A \textbf{72}, 012323 (2005)

\bibitem{fit05} Fitzsimons, J., Twamley, J.: Superballistic diffusion of entanglement in disordered spin chains. Phys. Rev. A \textbf{72}, 050301 (2005)

\bibitem{bur05} Burgarth, D., Bose, S.: Perfect quantum state transfer with randomly coupled quantum chains. New J. Phys. \textbf{7}, 135 (2005)

\bibitem{pet10} Petrosyan, D., Nikolopoulos, G. M., Lambropoulos, P.: State transfer in static and dynamic spin chains with disorder. 
Phys. Rev. A \textbf{81}, 042307 (2010)

\bibitem{zwi11} Zwick, A., \'Alvarez, G. A., Stolze, J., Osenda, O.: Robustness of spin-coupling distributions for perfect quantum state transfer. 
Phys. Rev. A \textbf{84}, 022311 (2011); Spin chains for robust state transfer: Modified boundary couplings versus completely engineered chains. 
\textbf{85}, 012318 (2012)

\bibitem{bru12} Bruderer, M., Franke, K., Ragg, S., Belzig, W., Obreschkow, D.: 
Exploiting boundary states of imperfect spin chains for high-fidelity state transfer. 
Phys. Rev. A \textbf{85}, 022312 (2012)

\bibitem{nik13} Nikolopoulos, G. M.: Characterizing quantum gates via randomized benchmarking. Phys. Rev. A \textbf{87}, 042311 (2013)

\bibitem{kur14} Zwick, A., \'Alvarez, G. A., Bensky, G., Kurizki, G.: Optimized dynamical control of state transfer through noisy spin chains. 
New. J. Phys. \textbf{16}, 065021 (2014)

\bibitem{ash15} Ashhab, S.: Quantum state transfer in a disordered one-dimensional lattice. Phys. Rev. A \textbf{92}, 062305 (2015)

\bibitem{pav16} Pavlis, A. K., Nikolopoulos, G. M., Lambropoulos, P.: Evaluation of the performance of two state-transfer Hamiltonians in the presence of static disorder. 
Quantum Inf Process \textbf{15}, 2553 (2016)

\bibitem{ron16} Ronke, R., Estarellas, M. P., D'Amico, I., Spiller, T. P., Miyadera, T.: Anderson localisation in spin chains for perfect state transfer. 
Eur. Phys. J. D \textbf{70}, 189 (2016)

\bibitem{lyr17} Almeida, G. M. A., de Moura, F. A. B. F., Apollaro, T. J. G., Lyra, M. L.: Disorder-assisted distribution of entanglement in XY spin chains. 
Phys. Rev. A \textbf{96}, 032315 (2017)

\bibitem{lyr17a} Almeida, G. M. A., de Moura, F. A. B. F., Lyra, M. L.: Quantum-state transfer through long-range correlated disordered channels. 
Phys. Lett. A \textbf{382}, 1335 (2018)

\bibitem{lyr17b} Almeida, G. M. A., de Moura, F. A. B. F., Lyra, M. L.: Entanglement generation between distant parties via disordered spin chains. 
Available from: arXiv:1711.08553 [quant-ph] (2017).

\bibitem{ben96} Bennett, C. H., DiVincenzo, D. P., Smolin, J. A., Wootters, W. K.: Mixed-state entanglement and quantum error correction. 
Phys. Rev. A \textbf{54}, 3824 (1996)

\bibitem{woo98} Wootters, W. K.: Entanglement of Formation of an Arbitrary State of Two Qubits. Phys. Rev. Lett. \textbf{80}, 2245 (1998)

\bibitem{vie18} Vieira, R., Rigolin, G.: Almost perfect transport of an entangled two-qubit state through a spin chain. Phys. Lett. A \textbf{382}, 2586 (2018)

\bibitem{vie13} Vieira, R., Amorim, E. P. M., Rigolin, G.: Dynamically Disordered Quantum Walk as a Maximal Entanglement Generator. 
Phys. Rev. Lett. \textbf{111}, 180503 (2013)

\bibitem{vie14} Vieira, R., Amorim, E. P. M., Rigolin, G.: Entangling power of disordered quantum walks.  Phys. Rev. A \textbf{89}, 042307 (2014)

\bibitem{dzy58} Dzyaloshinsky, I.: A thermodynamic theory of ``weak'' ferromagnetism of antiferromagnetics.  J. Phys. Chem. Solids \textbf{4}, 241 (1958)
		
\bibitem{mor60} Moriya, T: Anisotropic Superexchange Interaction and Weak Ferromagnetism. Phys. Rev. \textbf{120}, 91 (1960)

\end{thebibliography}
\end{document}